\documentclass{aastex631}

\shortauthors{Suarez and Moore}

\usepackage[graphicx]{realboxes}

\graphicspath{{./}{figures/}}

\begin{document}

\title{Estimations of Elemental Abundances During Solar Flares Observed in Soft X-rays by the MinXSS-1 CubeSat Mission}


\author[0000-0001-5243-7659]{Crisel Suarez}
\affiliation{Vanderbilt University\\
12201 West End Ave, Nashville, TN 37235} 
\affiliation{The Center for Astrophysics $\vert$ Harvard $\&$ Smithsonian\\
60 Garden St, Cambridge, MA 02138}

\author[0000-0002-4103-6101]{Christopher S. Moore}
\affiliation{The Center for Astrophysics $\vert$ Harvard $\&$ Smithsonian\\
60 Garden St, Cambridge, MA 02138}


\begin{abstract}
Solar flares are complex phenomena emitting all types of electromagnetic radiation, and accelerating particles on timescales of minutes, converting magnetic energy to thermal, radiative, and kinetic energy through magnetic reconnections. As a result, local plasma can be heated to temperatures in excess of 20 MK. During the soft X-ray (SXR) solar flare peak, the elemental abundance of low first-ionization potential (FIP) elements are typically observed to be depleted from coronal values. We explored the abundance variations using disk-integrated solar spectra from the Miniature X-ray Solar Spectrometer CubeSat-1 (MinXSS-1). MinXSS-1 is sensitive to the 1-12 keV energy range with an effective 0.25 keV full-width at half-maximum (FWHM) resolution at 5.9 keV. During the year-long mission of MinXSS-1, between May 2016 - May 2017, 21 flares with intermittent data downlinks were observed ranging from C to M class. We examine the time evolution of temperature, volume emission measure, and elemental abundances of Fe, Ca, Si, S, and Ar with CHIANTI spectral models near the peak SXR emission times observed in the MinXSS-1 data. We determined the average absolute abundance of A(Fe) = 7.81, A(Ca) = 6.84, A(S) = 7.28, A(Si) = 7.90, and A(Ar) = 6.56. These abundances are depleted from coronal values during the SXR peak compared to non-flaring times. The elemental abundance values that are depleted from their coronal values are consistent with the process of chromospheric evaporation, in which the lower atmospheric plasma fills the coronal loops.
\end{abstract}


\section{Introduction} \label{sec:intro}

The Miniature X-ray Solar Spectrometer (MinXSS) CubeSat-1 was the first solar science oriented CubeSat mission flown for the NASA Science Mission Directorate. Its primary role was to measure the solar soft X-ray (SXR) flux and determine its influence on Earth’s ionosphere and thermosphere \citep{woods2017minxss}. MinXSS-1 main onboard instruments consist of a spectrometer called X123 with an effective 0.25 keV full-width at half-maximum (FWHM) resolution at 5.9 keV and XP, a broadband X-ray photometer. {\color{Black}{These instruments are designed to detect photons from 0.5-30 keV with a programmed time integration time of 10s.}} Previous solar SXR observations have either been high spectral resolution with a narrow bandpass (Yohkoh, \cite{culhane1991braggSolarA} and Solar Maximum Mission (SMM), \cite{acton1980softSolarMax}), or spectrally integrated over a large bandpass (Geostationary Operational Environmental Satellites/X-ray Sensor (GOES/XRS), \cite{garcia1994temperaturegoes}). MinXSS-1 provides improved spectrally resolved solar SXR measurements over a broad bandpass which allows for the time evolution of plasma diagnostics including temperature, volume emission measure and, elemental abundances of thermal emission from solar flares \citep{woods2017minxss,Mason2016MinXSS, moore2018instrumentsminxss, mason2020minxss2, nagasawa2022minxssRhessi}. Similar broad bandpass SXR spectrometers include the MErcury Surface, Space ENvironment, GEochemistry, and Ranging/Solar Assembly for X-Rays (MESSENGER/SAX, \cite{solomon2001messenger, santo2001messenger, gold2001messenger}), Chandrayaan-1 and 2/X-ray Solar Monitor (XSM, \cite{goswami2009chandrayaan1, bhandari2005chandrayaan1, mithun2020solarCH2}), and CORONAS-PHOTON/Solar Photometer in X-rays (SphinX, \cite{Gburek2013Shpinx}).

The emission in the 1–10 keV energy range is {\color{Black}{dominated by free–free {\color{Black}{and}} free–bound thermal continuum radiation, but also includes bound-bound transitions forming spectral line complexes.}} Estimates of the abundances of different elements including Fe, Ca, Si, S, Ar, and Mg are derived from spectral fits to the measured count-rate spectra. This paper is primarily concerned with deriving estimates of the abundances of these elements for flares observed with MinXSS-1 and comparison with those from other solar SXR instruments.

Elemental abundances have been observed to play an important role in stars and planet formation \citep{maldonadoStellarPlanet2019}. Additionally, they can provide us with information about the plasma transport between the different layers of the solar atmosphere. Previous studies have observed that Fe, Ca, and Si, elements with low First Ionization Potential (FIP) (below ~10 eV), are 3-4 times more abundant in the corona relative to quiescent active regions (ARs) in the photosphere \citep{Feldman1992AbundancesUpper, feldman2003elementalFIP}. Similarly, \cite{delzanna2014elemental} found that the Fe/O and Fe/Ne ratios are normally increased by a factor of 3.2 compared to the photospheric values in quiescent active region cores. More recently, using multi-wavelength SXR and EUV observations of a quiescent AR \cite{delzanna2022multiwavelength}, showed that FIP elements are enhanced compared to their photospheric values. However, the inverse FIP bias, with the low FIP elements being depleted in the atmosphere relative to the high FIP elements has been observed near large sunspots producing large flares \citep{doschek2017sunspots}. {\color{Black}{Additionally, the emission during transient heating events in the transition region are consistent with {\color{Black}{photospheric composition}} \citep{warren2016transitionImpulsive}}}. Similarly, active stars have shown a similar pattern of the inverse FIP bias \citep{laming2015fipsolarStellar}.

There are several models proposed that explain the observed enhancement in the corona \citep{henoux1998fiptheory, feldman2003elementalFIP, laming2004unifiedFIP, laming2009non, laming2021fip}. Most of the models consist of the fractionation of the low-FIP ions from neutral atoms in the chromosphere which rise into the corona. \cite{laming2004unifiedFIP, laming2009non, laming2021fip} describes this phenomenon with the ponderomotive force arising from the propagation of Alfvén waves through the chromosphere. The amount of elemental enhancement in the corona depends on the amplitude and frequency of the Alfvén waves relative to the coronal flux loop lengths and wave flux. 

In this paper, we report on the analysis of 21 well- observed MinXSS-1 solar flares during May 2016-2017. Abundance estimates during the peak of soft X-ray flux for the elements Fe, Ca, S, Si, Ar, and Mg are given. 
The observations and analysis procedures are described in the following sections and compared to previous results of elemental abundance during flares. 

\section{OBSERVATIONS AND DATA ANALYSIS} \label{sec:style}

The main objective of our analysis is to study the evolution of plasma temperature, volume emission measure, and elemental abundance during flares that were observed by MinXSS-1. The observed SXR spectrum, is dominated by the continuum and line complexes of ionized Fe near 1.1 keV and 6.7 keV, Mg near 1.3 and 1.5 keV, Si near 1.8 and 2.1 keV, S near 2.7 keV, Ar near 3 keV, Ca near 4 keV, and the Fe+Ni complex near 8 keV. The slope of the spectrum is dependent on the temperature and the emission measure determines the absolute photon flux of the spectra. The observed count rate spectra were fitted using series of parametric spectra fits that are available in the OSPEX (\cite{2020TolbertSchwartzOSPEX}, \textbf{\url{https://hesperia.gsfc.nasa.gov/ssw/packages/spex/doc/ospex_explanation.htm}}) programming suite. OSPEX is available in Solar Software, which utilizes the Chianti Atomic Database version 9 \citep{dere1997chianti, Dere2019Chianti} and fitting routines which automatically adjust the best fit parameters by improving the goodness of fit determined by the reduced chi-squared statistic. 

MinXSS-1 {\color{Black}{observations were}} nominally collected with a cadence of 10 second, however we average the data over 1 minute intervals.  {\color{Black}{We selected a low energy value for the fitting procedure of 1.1 keV due to the detector sensitivity and uncertainties with the instrument response below 1 keV. These uncertainties are related to the efficiency of the detector and photoelectric interactions of the incident radiation with the detector below 1 keV. These processes are further explained in \citep{moore2018instrumentsminxss} }.} The high energy value of the spectra fitting energy range {\color{Black}{varies with time and is determined automatically by searching the energy bin where the average count rate spectra is greater or equal to  2 counts per integration time.  Therefore, the high energy limit has at least 2 counts. }} MinXSS-1 observed $\sim$ 30 C and M class flares in the May 2016-2017 mission. A complete catalog of all MinXSS-1 flares is available at \url{https://lasp.colorado.edu/home/minxss/data/}. For this study we only consider flares that had coverage at the SXR peak of the flare, reducing our sample to 21 flares. We consider the peak of the flares as it provides the highest signal in SXR. A full catalog of the 21 flares studied and their best fit parameters for temperature, volume emission measure, and elemental abundance are presented in Table \ref{tbl:BKD_Params} during non-flaring (background) times and Table \ref{tbl:Flare_Params} during the peak flaring time.


In this study we used two sets of two-variable thermals with separate abundances (2vth$\_$abun in OSPEX) with a spectral range of 1.1–12 keV. Each set of two-variable thermal fits can autonomously adjust the volume emission measure, temperature, and the abundances of Fe, Ca, S, Si, Ar, and Mg. The individual elemental abundances are only adjusted if the average signal-to-noise count rate of each element is higher than the signal-to-noise ratio threshold (S/N = 9). {\color{Black}{For this study, we calculate the average signal-to-noise of ten 0.03 keV energy bins around each element's spectral feature. If the averaged signal-to-noise ratio is greater than 9, the abundance parameters are set as free parameters in the spectra fit. Otherwise, we assume a coronal abundance for each abundance parameter.}} Additionally, {\color{Black}{due to detector effects and variations in the spacecraft’s orbital environment we fit the energy gain and offset of the spectra as a free parameter for our spectra fits. See \citep{moore2018instrumentsminxss} for an explanation of the MinXSS-X123 detector. } }

\begin{figure}[ht!]
\plotone{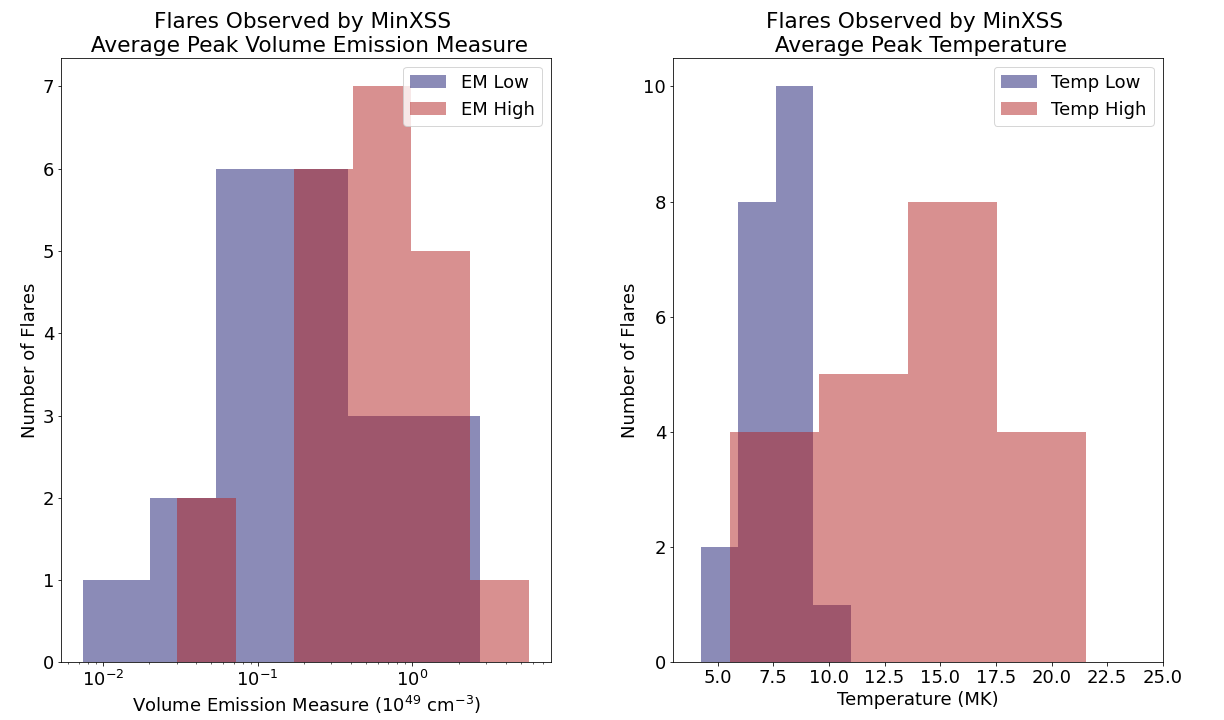}
\caption{Histogram of the average peak volume emission measure and temperature of 21 flares observed by MinXSS-1. The volume emission measure ranges from $10^{47} - 10^{49}$ cm$^{-3}$ and temperatures from $\sim$4 to  21 MK.}
\label{fig:hist_temp_em}
\end{figure}

\begin{figure}[ht!]
\plotone{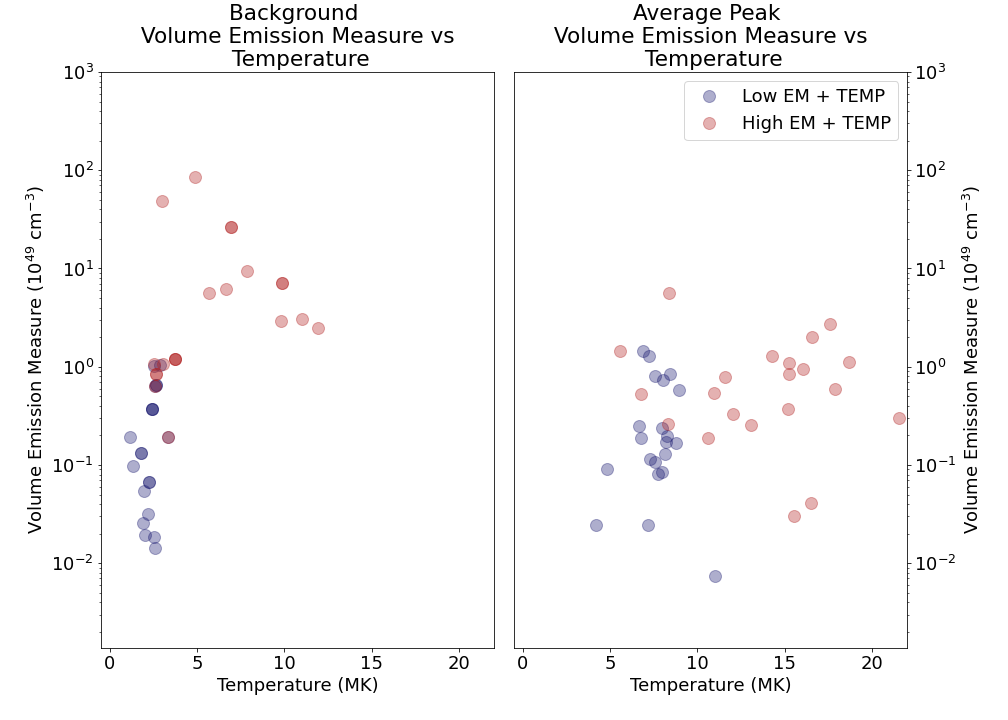}
\caption{Relationship of volume emission measure and temperature during non-flaring (background) and SXR peaking times. The values are represented in Tables \ref{tbl:BKD_Params} and \ref{tbl:Flare_Params}. In the background panel (left), the lower temperature corresponds to a lower volume emission measure. The darker circles are where a both the low and high temperatures converge to a single value. In the average SXR peak panel (right), there is no clear relationship between volume emission measure and temperature. }
\label{fig:scatterEMvsTEMP}
\end{figure}

\begin{figure}[ht!]
\plotone{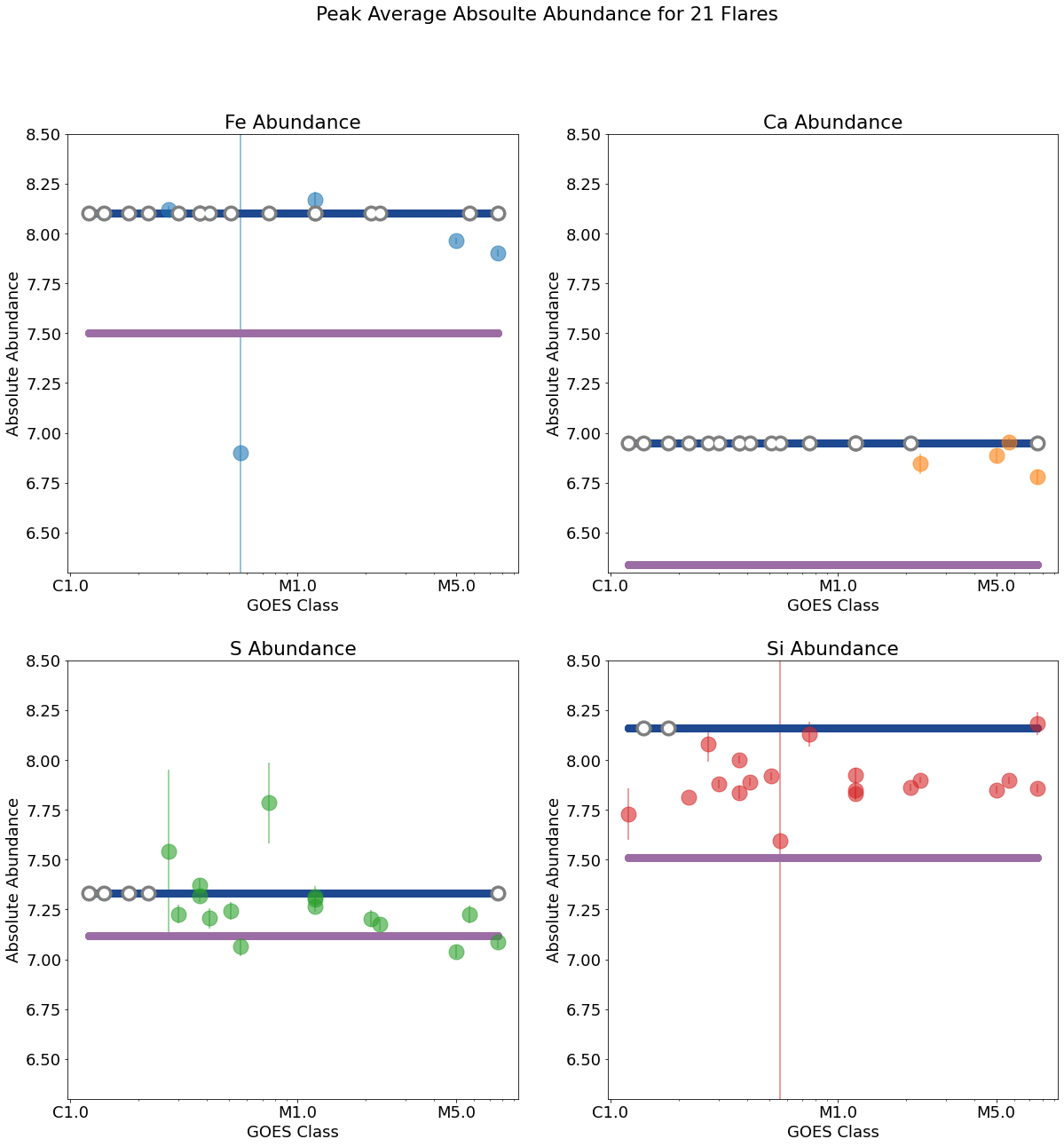}
\caption{Average absolute abundance during the SXR peak for 21 flares observed by MinXSS-1. The white circles are where the elemental abundance parameter are set to coronal values, while the color circles are fitted parameters. The dark blue line shows the coronal abundances from  \cite{feldman2003elementalFIP} and the purple line shows the photospheric abundances from \cite{asplund2009chemical}. {\color{Black}{The absolute abundances are on logarithmic scale with A(H) = 12. We note that most fitted parameters are {\color{Black}{below}} the coronal abundance implying that  the lower  atmospheric plasma is filling the coronal loops. The Fe abundance parameter is only fitted for the 6.7 keV feature. The Fe abundance for the C5.6 flare has large uncertainties due to a high abundance value during the non-flaring time period.}}  }. 
 \label{fig:all_flares_abund_multi}
\end{figure}

For the first two-variable thermal fit, we averaged at least three minutes of non-flaring (background) data. The times used are denoted in Table \ref{tbl:BKD_Params}. Due to the irregular intervals of the MinXSS-1, some time intervals are several hours. However, there might only be three minutes of MinXSS-1 spectra. This non-flaring spectra fit is used as an enhancement when fitting the spectra of the flare. {\color{Black}{We used this approach instead of subtracting the background since it is likely that during a flare there are contributions from non-flaring plasma in the observed spectra.}} Table \ref{tbl:BKD_Params} shows the two volume emission measures, two temperatures, and elemental abundances during non-flaring times. Table \ref{tbl:Flare_Params} shows the averaged values of the two volume emission measures, two temperatures, and elemental abundances during the SXR peak of all 21 flares.  Figure \ref{fig:hist_temp_em} shows the distribution of volume emission measures ranging from $10^{47}$ to $10^{50}$ cm$^{-3}$ and temperatures from 4 to 21 MK, for all 21 flares observed by MinXSS-1. Figure \ref{fig:scatterEMvsTEMP} indicates no clear relationship between the volume emission measure and temperature values during non-flaring and flaring times. Figure \ref{fig:all_flares_abund_multi} shows the average abundance values of Fe, Ca, Si, and S during the SXR peak for all 21 flares compared to \cite{feldman2003elementalFIP} and \cite{asplund2009chemical} values.

\begin{longrotatetable}
     \begin{deluxetable*}
{|l|l|l|l|l|l|l|l|l|l|l|l|l|}
 \tablecaption{Non-flaring parameters used as a fixed component for each flare. The Start BKD and End BKD Times denote the time interval when the spectra was averaged. Due to the intermittence data from MinXSS-1, some time intervals are several hours. However, there might only be a couple of minutes of MinXSS-1 spectra. The parameters from the background volume emission measure (VEM, $\times$ 10$^{49}$ cm$^{-3}$), temperature (Temp, MK), and abundances. The absolute abundances are on a logarithmic scale with A(H) = 12. If the abundances are observed to have low signal, we assume the coronal values from  \cite{feldman2003elementalFIP}. As a reference the \cite{feldman2003elementalFIP} values are: Fe (8.10), Ca (6.95), S (7.33), Si (8.16), Mg (8.18), and Ar (6.59). The arithmetic mean takes into account the average peak values for all flares,  the weighted arithmetic mean only takes into account the flares where the fitted parameters have a signal-to-noise ratio greater than nine. $^a$Volume Emission Measure (VEM $\times$ 10$^{49}$ cm$^{-3}$), $^b$Temperature (MK), $^*$Fitted parameters with signal-to-noise ratio greater than nine (S/N $>$ 9). \label{tbl:BKD_Params}}
\tablewidth{0pt}
\tabletypesize{\footnotesize}
\tablehead{
\colhead{Start BKD Time} &
\colhead{End BKD Time}  & 
\colhead{Class} &  
\colhead{Low} &
\colhead{High} &
\colhead{Low} &
\colhead{High} &
\colhead{Fe} &
\colhead{Ca} &
\colhead{S} &
\colhead{Si} &
\colhead{Mg} &
\colhead{Ar} \\
\colhead{} &
\colhead{} &
\colhead{} &
\colhead{VEM$^a$} &
\colhead{VEM$^a$} & 
\colhead{Temp$^b$ } &
\colhead{Temp$^b$ } 
}     
    \startdata{
    2016 Jun 13 1:50:30 & 2016 Jun 13 5:32:30 & C3.0 & 0.10 & 86.05 & 1.32 & 4.87 & 8.10 & 6.95 & 7.33 & 8.16 & 8.18 & 6.59 \\ \hline
    2016 Jul 07 7:06:30 & 2016 Jul 07 7:48:30 & C5.1 & 0.03 & 5.63 & 1.93 & 5.67 & 8.10 & 6.95 & 7.33 & 8.16 & 7.99$^*$ & 6.59 \\ \hline
    2016 Jul 07 23:55:00 & 2016 Jul 08 0:47:00 & C2.7 & 0.02 & 6.14 & 2.03 & 6.66 & 8.10 & 6.95 & 7.33 & 8.16 & 7.99$^*$ & 6.59 \\ \hline
    2016 Jul 08 1:40:00 & 2016 Jul 19 9:40:00 & C2.2 & 1.02 & 1.07 & 2.52 & 2.54 & 8.10 & 6.95 & 7.33 & 8.16 & 8.18 & 6.59 \\ \hline
    2016 Jul 21 0:06:00 & 2016 Jul 21 0:12:30 & M1.2 \& C3.7 & 0.07 & 7.09 & 2.24 & 9.84 & 8.10 & 6.95 & 7.33 & 8.16 & 8.18 & 6.59 \\ \hline
    2016 Jul 21 0:06:00 & 2016 Jul 21 0:12:30 & M1.0 & 0.07 & 7.09 & 2.24 & 9.84 & 8.10 & 6.95 & 7.33 & 8.16 & 8.18 & 6.59 \\ \hline
    2016 Jul 21 15:16:00 & 2016 Jul 22 6:15:00 & C1.2 & 0.02 & 2.49 & 2.52 & 11.93 & 8.10 & 6.95 & 7.33 & 8.16 & 8.18 & 6.59 \\ \hline
    2016 Jul 23 1:39:00 & 2016 Jul 23 1:42:30 & M5.0 & 0.37 & 1.19 & 2.45 & 3.76 & 8.10 & 6.95 & 7.33 & 7.85$^*$ & 7.98$^*$ & 6.59 \\ \hline
    2016 Jun 12 19:47:00 & 2016 Jun 12 20:29:00 & C1.8 & 0.01 & 3.03 & 2.58 & 11.02 & 8.10 & 6.95 & 7.33 & 8.16 & 8.05$^*$ & 6.59 \\ \hline
    2016 Jul 23 1:39:00 & 2016 Jul 23 1:42:30 & C1.4 & 0.37 & 1.19 & 2.45 & 3.76 & 8.10 & 6.95 & 7.33 & 7.85$^*$ & 7.98$^*$ & 6.59 \\ \hline
    2016 Jul 23 1:39:00 & 2016 Jul 23 1:42:30 & M7.6  \& M5.5 & 0.37 & 1.19 & 2.45 & 3.76 & 8.10 & 6.95 & 7.33 & 7.85$^*$ & 7.98$^*$ & 6.59 \\ \hline
    2016 Sep 21 11:34:00 & 2016 Sep 21 11:39:30 & C1.7 & 0.05 & 9.33 & 1.99 & 7.87 & 8.10 & 6.95 & 7.33 & 8.16 & 8.14$^*$ & 6.59 \\ \hline
    2016 Sep 22 5:33:00 & 2016 Sep 22 5:36:30 & C5.6 & 0.19 & 0.19 & 3.33 & 3.33 & 9.02$^*$ & 6.95 & 7.33 & 8.16 & 8.18 & 6.59 \\ \hline
    2016 Nov 29 15:50:00 & 2016 Nov 29 16:35:00 & M1.2 \& C1.9 & 0.03 & 2.89 & 2.21 & 9.83 & 8.10 & 6.95 & 7.33 & 8.16 & 8.16$^*$ & 6.59 \\ \hline
    2016 Nov 29 3:51:00 & 2016 Nov 29 6:47:00 & C7.5 & 0.63 & 0.63 & 2.59 & 2.60 & 8.10 & 6.95 & 7.33 & 8.16 & 8.18 & 6.59 \\ \hline
    2017 Feb 22 12:50:30 & 2017 Feb 22 13:02:30 & C4.1 & 0.19 & 48.29 & 1.17 & 2.99 & 8.10 & 6.95 & 7.33 & 8.16 & 8.18 & 6.59 \\ \hline
    2017 Apr 01 16:57:00 & 2017 Apr 01 19:29:00 & C3.7 & 0.65 & 0.85 & 2.65 & 2.67 & 8.10 & 6.95 & 7.33 & 8.16 & 8.18 & 6.59 \\ \hline
    2017 Apr 01 16:57:00 & 2017 Apr 01 19:29:00 & M4.4 & 0.65 & 0.85 & 2.65 & 2.67 & 8.10 & 6.95 & 7.33 & 8.16 & 8.18 & 6.59 \\ \hline
    2017 Apr 02 6:37:00 & 2017 Apr 02 7:49:00 & M2.3 & 1.05 & 1.07 & 2.89 & 3.08 & 8.10 & 6.95 & 7.33 & 8.16 & 8.04$^*$ & 6.59 \\ \hline
    2017 Apr 02 16:11:01 & 2017 Apr 02 17:54:30 & M2.1 & 0.13 & 26.57 & 1.78 & 6.95 & 8.10 & 6.95 & 7.33 & 8.16 & 8.15$^*$ & 6.59 \\ \hline
    2017 Apr 02 16:11:01 & 2017 Apr 02 17:54:30 & M5.7 & 0.13 & 26.57 & 1.78 & 6.95 & 8.10 & 6.95 & 7.33 & 8.16 & 8.15$^*$ & 6.59 \\ \hline
     & ~ & ~ & ~ & ~ & ~ & ~ & ~ & ~ & ~ & ~ & ~ & ~  \\ \hline
   \textbf{Arithmetic Mean} & ~ &~ & 0.29 & 11.40 & 2.27 & 5.84 & 8.14 & 6.95 & 7.33 & 8.12 & 8.12 & 6.59 \\ \hline
    \textbf{Weighted Arithmetic Mean} & ~ &~ & ~ & ~ & ~ & ~ & 9.02 & ~ & ~ & 7.85 & 8.07 & ~ }\enddata
\end{deluxetable*}
\end{longrotatetable}


\begin{longrotatetable}
\begin{deluxetable*}{|l|l|l|l|l|l|l|l|l|l|l|l|l|}
 \tablecaption{Average SXR peak volume emission measure (VEM $\times$ 10$^{49}$ cm$^{-3}$), temperature (MK), and elemental abundances. The peak time is approximately the SXR peaking time. We report the averaged parameters during 10 minutes of SXR peaking time. The absolute abundances are on a logarithmic scale with A(H) = 12. If the abundances are observed to have low signal, we assume the coronal values from  \cite{feldman2003elementalFIP}. As a reference the \cite{feldman2003elementalFIP} values are: Fe (8.10), Ca (6.95), S (7.33), Si (8.16), Mg (8.18), and Ar (6.59). The arithmetic mean takes into account the average peak values for all flares, the weighted arithmetic mean only takes into account the flares where the fitted parameters have a signal-to-noise ratio greater than nine. \\ $^*$Fitted parameters with signal-to-noise ratio greater than nine (S/N $>$ 9). \label{tbl:Flare_Params}}
\tablewidth{0pt}
\tabletypesize{\footnotesize}
\tablehead{
\colhead{Date} &
\colhead{Peak Time}  & 
\colhead{Class} &  
\colhead{Low EM } &
\colhead{High EM} & 
\colhead{Low Temp} &
\colhead{High Temp} &
\colhead{Fe} &
\colhead{Ca} &
\colhead{S} &
\colhead{Si} &
\colhead{Mg} &
\colhead{Ar} \\
\colhead{} &
\colhead{} &
\colhead{} &
\colhead{ (10$^{49}$ cm$^{-3}$ )} &
\colhead{ (10$^{49}$ cm$^{-3}$ )} &
\colhead{ (MK)} &
\colhead{ (MK)} 
}   
    \startdata{
    2016 Jun 13 & 05:52:00 & C3.0 & 0.11 & 0.19 & 7.58 & 10.63 & 8.10 & 6.95 & 7.23$^*$ & 7.88$^*$ & 8.40$^*$ & 6.59 \\ \hline
    2016 Jul 07 & 07:56:00 & C5.1 & 0.09 & 0.37 & 7.99 & 15.16 & 8.10 & 6.95 & 7.24$^*$ & 7.92$^*$ & 8.29$^*$ & 6.59 \\ \hline
    2016 Jul 08 & 00:56:00 & C2.7 & 0.11 & 0.33 & 7.30 & 12.06 & 8.12$^*$ & 6.95 & 7.54$^*$ & 8.08$^*$ & 8.34$^*$ & 6.57$^*$ \\ \hline
    2016 Jul 19 & 11:53:00 & C2.2 & 0.19 & 0.53 & 6.77 & 6.79 & 8.10 & 6.95 & 7.33 & 7.81$^*$ & 8.27$^*$ & 6.59 \\ \hline
    2016 Jul 21 & 00:46:00 & M1.2 \& C3.7 & 0.24 & 0.26 & 7.99 & 13.05 & 8.10 & 6.95 & 7.30$^*$ & 7.85$^*$ & 8.31$^*$ & 6.59 \\ \hline
    2016 Jul 21 & 01:49:00 & M1.0 & 0.08 & 0.59 & 7.76 & 17.87 & 8.17$^*$ & 6.95 & 7.26$^*$ & 7.92$^*$ & 8.34$^*$ & 6.43$^*$ \\ \hline
    2016 Jul 22 & 07:26:00 & C1.2 & 0.09 & 1.46 & 4.83 & 5.56 & 8.10 & 6.95 & 7.33 & 7.73$^*$ & 7.94$^*$ & 6.59 \\ \hline
    2016 Jul 23 & 02:11:00 & M5.0 & 1.29 & 2.00 & 7.24 & 16.56 & 7.96$^*$ & 6.89$^*$ & 7.04$^*$ & 7.85$^*$ & 8.22$^*$ & 6.31$^*$ \\ \hline
    2016 Jul 23 & 12:57:00 & C1.8 & 0.02 & 0.03 & 7.20 & 15.53 & 8.10 & 6.95 & 7.33 & 8.16 & 8.51$^*$ & 6.59 \\ \hline
    2016 Jul 23 & 03:42:00 & C1.4 & 0.02 & 0.78 & 4.23 & 11.59 & 8.10 & 6.95 & 7.33 & 8.16 & 8.31$^*$ & 6.59 \\ \hline
    2016 Jul 23 & 05:16:00 & M7.6 \& M5.5 & 1.46 & 2.74 & 6.87 & 17.61 & 7.90$^*$ & 6.78$^*$ & 7.09$^*$ & 7.86$^*$ & 8.13$^*$ & 6.20$^*$ \\ \hline
    2016 Sep 21 & 11:51:00 & C1.7 & 0.01 & 0.04 & 10.99 & 16.52 & 8.10 & 6.95 & 7.33 & 8.18$^*$ & 8.80$^*$ & 6.59 \\ \hline
    2016 Sep 22 & 05:47:00 & C5.6 & 0.13 & 0.30 & 8.14 & 21.53 & 6.90$^*$ & 6.95 & 7.07$^*$ & 7.59$^*$ & 8.20$^*$ & 6.56$^*$ \\ \hline
    2016 Nov 29 & 23:38:00 & M1.2 \& C1.9 & 0.17 & 1.09 & 8.80 & 15.23 & 8.10 & 6.95 & 7.31$^*$ & 7.83$^*$ & 8.38$^*$ & 6.59$^*$ \\ \hline
    2016 Nov 29 & 07:10:00 & C7.5 & 0.20 & 0.55 & 8.25 & 10.96 & 8.10 & 6.95 & 7.79$^*$ & 8.13$^*$ & 8.48$^*$ & 6.89$^*$ \\ \hline
    2017 Feb 22 & 13:27:00 & C4.1 & 0.25 & 0.26 & 6.65 & 8.33 & 8.10 & 6.95 & 7.21$^*$ & 7.89$^*$ & 8.26$^*$ & 6.59 \\ \hline
    2017 Apr 01 & 19:56:00 & C3.7 & 0.17 & 5.62 & 8.20 & 8.38 & 8.10 & 6.95 & 7.32$^*$ & 7.84$^*$ & 8.35$^*$ & 6.59 \\ \hline
    2017 Apr 01 & 21:48:00 & M4.4 & 0.80 & 1.30 & 7.56 & 14.27 & 8.10 & 6.95 & 7.37$^*$ & 8.00$^*$ & 8.46$^*$ & 6.82$^*$ \\ \hline
    2017 Apr 02 & 13:00:00 & M2.3 & 0.74 & 0.84 & 8.02 & 15.25 & 8.10 & 6.85$^*$ & 7.17$^*$ & 7.90$^*$ & 8.41$^*$ & 6.56$^*$ \\ \hline
    2017 Apr 02 & 18:38:00 & M2.1 & 0.58 & 0.95 & 8.94 & 16.01 & 8.10 & 6.95 & 7.20$^*$ & 7.86$^*$ & 8.38$^*$ & 6.65$^*$ \\ \hline
    2017 Apr 02 & 20:33:00 & M5.7 & 0.84 & 1.11 & 8.43 & 18.66 & 8.10 & 6.95$^*$ & 7.22$^*$ & 7.90$^*$ & 8.36$^*$ & 6.56$^*$ \\ \hline
     ~ & ~ & ~ & ~ & ~ & ~ & ~ & ~ & ~ & ~ & ~ & ~ & ~ \\ \hline 
    \textbf{Arithmetic Mean} & ~ & ~ & 0.36 & 1.02 & 7.61 & 13.69 & 8.03 & 6.93 & 7.29 & 7.92 & 8.34 & 6.57 \\ \hline
      \textbf{Weighted Arithmetic Mean} & ~ & ~ & ~ & ~ & ~ & ~ & 7.81 & 6.84 & 7.28 & 7.90 & 8.34 & 6.56 }\enddata
   \end{deluxetable*}
   \end{longrotatetable}

\section{Results}

\subsection{Spectral results of C4.1, M1.0, M5.0 and M7.7 flares } 

\begin{figure}[ht!]
\begin{interactive}{animation}{Videos/AIA_CONTEXT_VIDEO_2.mp4}
\plotone{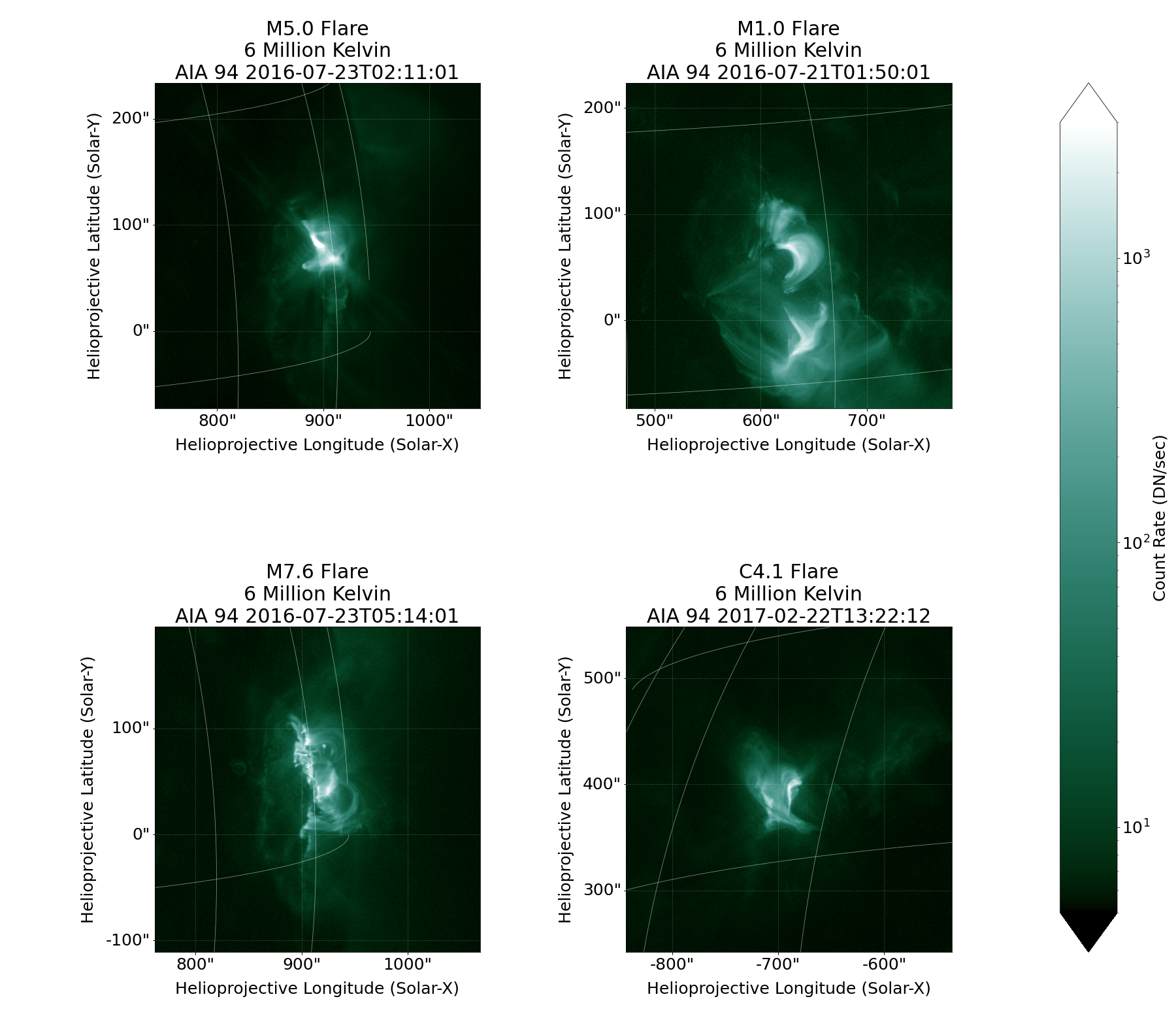}
\end{interactive}
\caption{{SDO/AIA 94 Å context images for the C4.1, M1.0, M5.0, and M7.7 flares. The time shown in each image corresponds to approximately the peak of each flare. The field of view of each image is 512x512. A video of the spatial evolution of the coronal arcade loops during the impulsive, peak and decay phases of the four flares is available in the online version of this paper (Video duration: 00:07).}}
\label{fig:aia_flares}
\end{figure}

Detailed spectral results of four flares observed by MinXSS-1 are given as an illustration of the parameters obtained for all 21 flares that were observed over the year-long mission. Figure \ref{fig:aia_flares} shows the Solar Dynamics Observatory/Atmospheric Imaging Assembly (SDO/AIA) 94 Å context images at the peak of four different flares with GOES Class M5.0, M1.0, M7.7, and C4.1. The images probe a temperature of approximately 6 million Kelvin \citep{lemen2012atmospheric}. A video of the four flares is available in the online version of this paper.

\begin{figure}[ht!]
\includegraphics[scale=0.2]{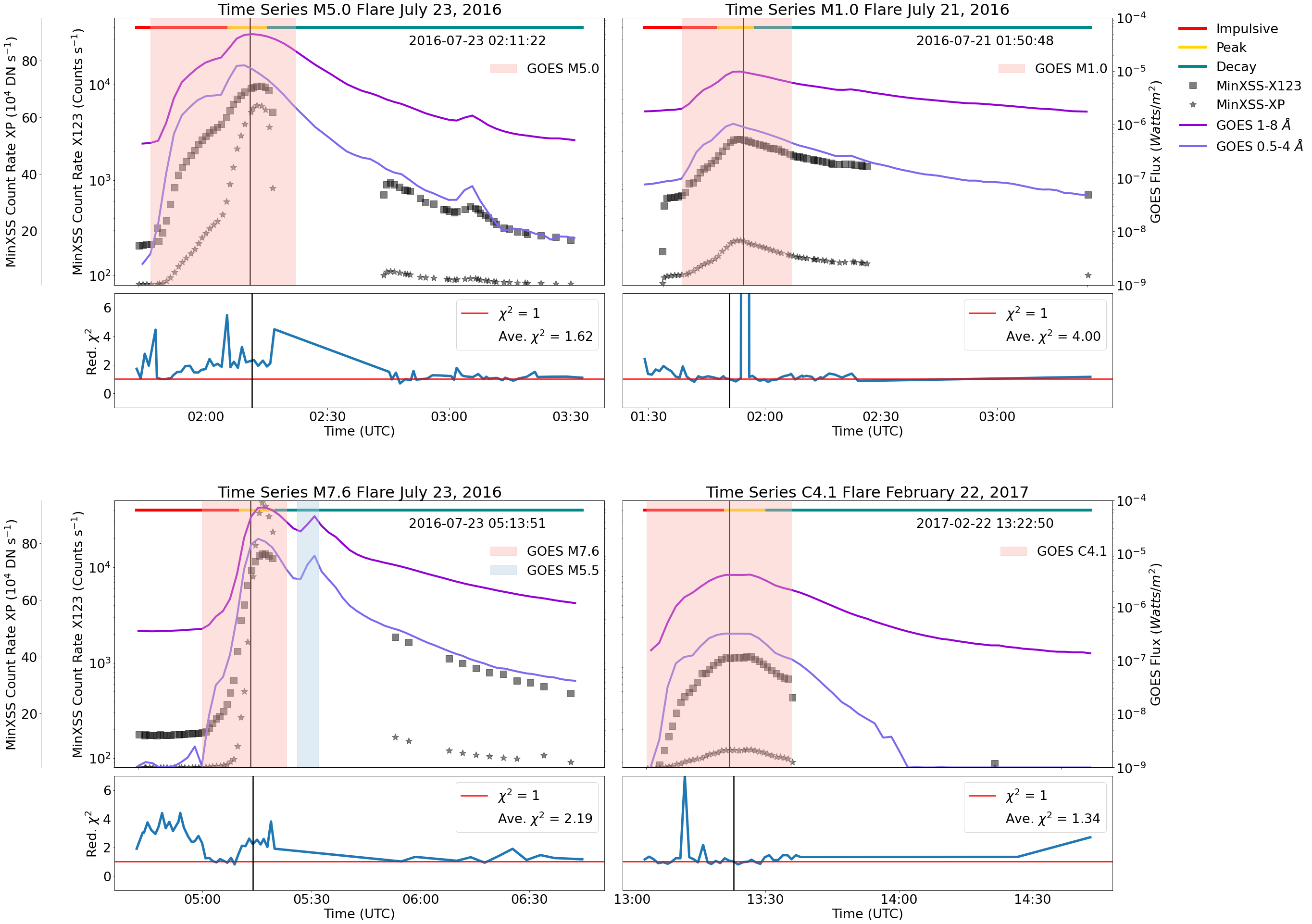}

\caption{Time Series plot for four flares observed by MinXSS-1. These four flares will illustrate the analysis and process of all the flares MinXSS-1 observed. The top left is the July 23, 2016 M5.0 flare, top right the July 21, 2016 M1.0 flare, bottom left is the July 23, 2016 M7.6 and bottom right is the C4.1 February 22, 2017. The pink and blue shaded region are the GOES predefined start and end times. The MinXSS-X123 is shown in black boxes, MinXSS-XP in black stars and GOES/XRS 1-8 \AA in purple, 0.5-4 \AA in blue. On top of each panel the red, yellow, and green bars show the impulsive, peak and decay phase of each flare. Additionally, on the bottom of each flare we show the evolution of reduced chi-square from the spectra fit as a function of time. We note that during the peak of the flare the reduced chi-square value is approximately 1.  \label{fig:timeseries_minxss}}
\end{figure}

\begin{figure}[ht!]
\begin{interactive}{animation}{Videos/Specrta_Quad_video_test_diff_line_fast.mp4}
\plotone{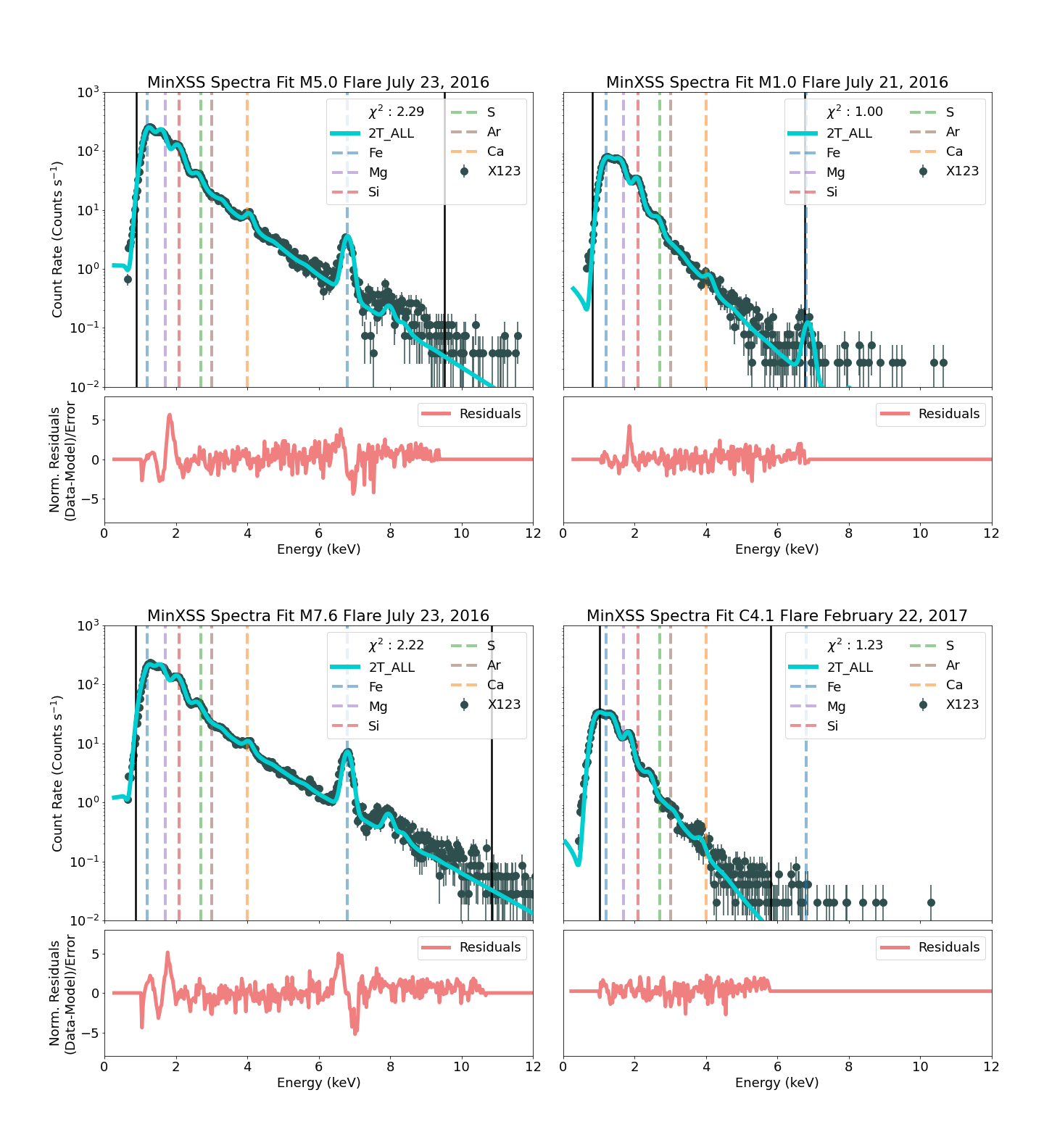}]
\end{interactive}
\caption{{Count rate spectra fit during the peak SXR with normalized residuals for four flares observed by MinXSS-1 X123. The top left is the July 23, 2016 M5.0 flare, top right the July 21, 2016 M1.0 flare, bottom left is the July 23, 2016 M7.6 and bottom right is the C4.1 February 22, 2017. The SXR peak times of these four flares are indicated by the vertical black line in Figure \ref{fig:timeseries_minxss}. The observed MinXSS-1 X123 spectra is shown in dark gray, and the double emission measure and temperature spectra fit in cyan. The black vertical lines represent the low and high energy limits for the fit. The colored vertical lines approximately represent the elemental features for Fe (1.2 and 6.7 keV), Ca (4.0 keV), Mg (1.7 keV), Si (2.1 keV), S (2.7 keV), and Ar (3.0 keV). A video showing the temporal evolution of SXR spectra at the impulsive, peak and decay phases of the four flares is available in the online version of this paper. In the video, the slope of the spectrum changes with respect to changes in temperature and volume emission measure. Additionally, the elemental abundances features change during the flare. (Video duration: 00:05).}} 
\label{fig:Spectra_Minxss}
\end{figure}

Figure \ref{fig:timeseries_minxss} shows the SXR light curves observed by MinXSS-1 X123, MinXSS-1 XP, and GOES/XRS 1-8 Å and 0.5–4 Å. The GOES/XRS light curves of each of the flares show a gradual impulsive phase followed by the SXR peak and then a gradual decay. These phases are shown on top of each plot and color-coded red, yellow, and green respectively. The black line denotes the approximate SXR peak and the time of the flare spectra in Figure \ref{fig:Spectra_Minxss}.

We describe the fitting procedure for the spectrum taken during each flare. The observed spectrum, is dominated by {\color{Black}{the free–free, and free–bound thermal continuum radiation}}, with line complexes of ionized Fe near 1.1 keV and 6.7 keV, Mg near 1.3 and 1.5 keV, Si near 1.8 and 2.1 keV, S near 2.7 keV, Ar near 3 keV, Ca near 4 keV, and the Fe+Ni complex near 8 keV. The slope of the spectrum is dependent on the temperature and the emission measure determines the absolute photon flux of the spectra. The observed count rate spectra were fitted with a series of parametric fits using the OSPEX and the Chianti Atomic Database version 9 (\cite{2020TolbertSchwartzOSPEX}, \textbf{\url{https://hesperia.gsfc.nasa.gov/ssw/packages/spex/doc/ospex_explanation.htm}}, \cite{dere1997chianti, Dere2019Chianti}). We assumed a two temperature (with two values for temperature and two values for volume emission measure) isothermal emitting region for the 1.1–12 keV spectral range and let the energy gain, energy offset, volume emission measure, temperature, and the abundances of Fe, Ca, S, Si, Ar, and Mg be free parameters and adjusted as necessary in the fitting procedure.  To model the spectra, we first modeled the spectrum of the non-flaring plasma (background) independently and included it as a fixed component while fitting the flare spectrum. {\color{Black}{Therefore, our spectra results are linear combinations of non-flaring and flaring fits.  We opted for this approach as the enhanced spectra method had an average reduced chi-square of approximately 1. Moreover, when subtracting the background, the uncertainties of our spectra results increase,  have higher residuals at the lower energies, and higher reduced chi-square than the enhanced spectra (see Figure \ref{fig:enhanced_subtracted} for comparison).
}}

\begin{figure}[ht!]
\plotone{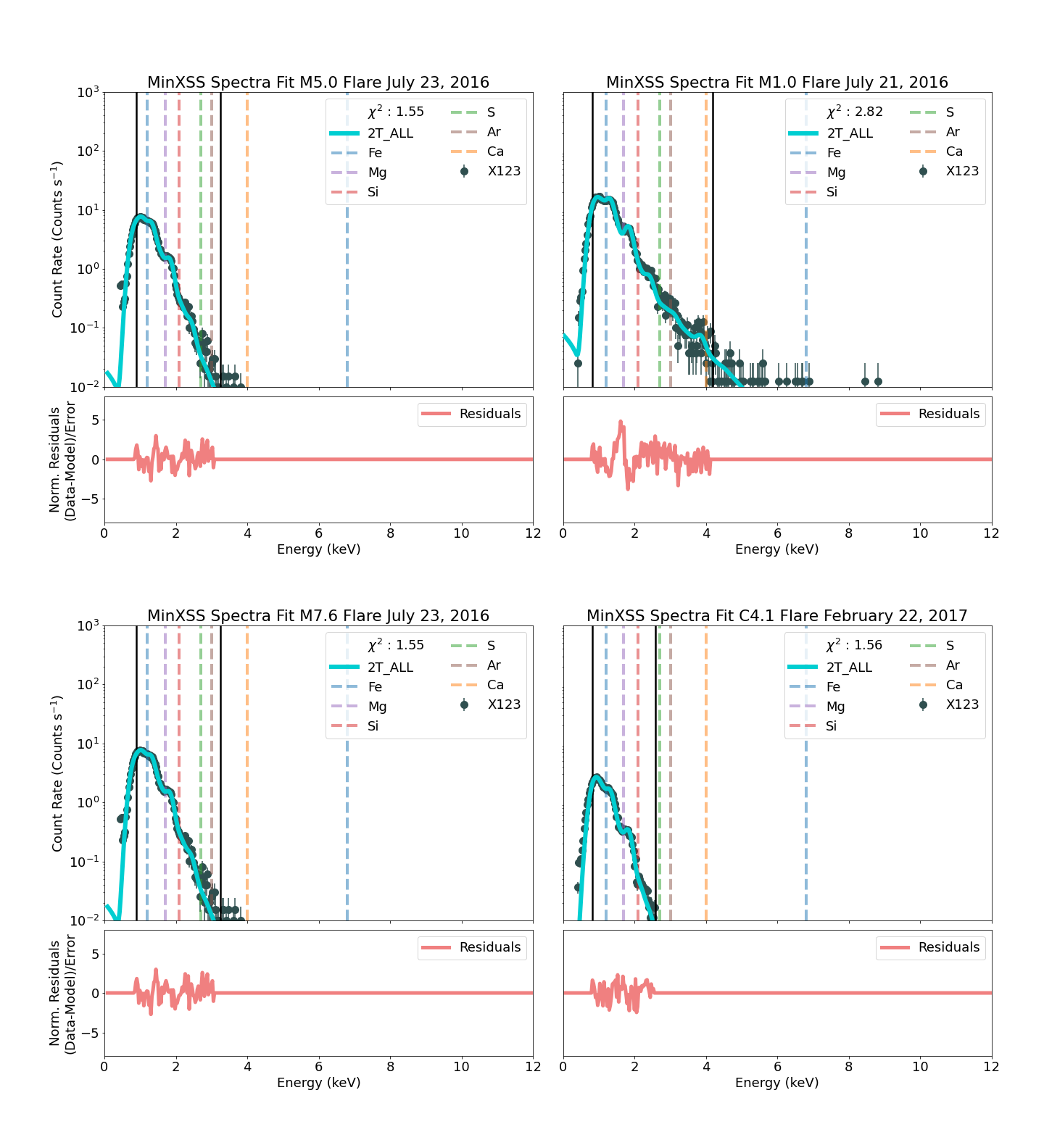}
\caption{Count rate spectra fit during the non-flaring (background) times with normalized residuals for four flares observed by MinXSS-1 X123. The top left is the July 23, 2016 M5.0 flare, top right the July 21, 2016 M1.0 flare, bottom left is the July 23, 2016 M7.6 and bottom right is the C4.1 February 22, 2017. The observed MinXSS-1 X123 spectra is shown in dark gray, and the double emission measure and temperature spectra fit in cyan. The black vertical lines represent the low and high energy limits for the fit. The colored vertical lines approximately represent the elemental features for Fe (1.2 and 6.7 keV), Ca (4.0 keV), Mg (1.7 keV), Si (2.1 keV), S (2.7 keV), and Ar (3.0 keV). } 
 \label{fig:Spectra_Minxss_BKD}
\end{figure}

An example of the fixed non-flaring component of the four flares is shown in Figure \ref{fig:Spectra_Minxss_BKD}. The fixed non-flaring spectra represents the quiescent solar conditions before the flare occurs. The flare spectrum represents the flaring event and the change of solar conditions. We analyzed both components since MinXSS-1 integrates over the whole Sun. 

For the non-flaring fixed spectra, we averaged at least three minutes of non-flaring data before the flare occurred to increase the statistical significance of the background. Figure \ref{fig:Spectra_Minxss} shows the count rate spectra fitting results at the SXR peak during the four flares.  {\color{Black}{A video of the spectra fit for the whole flare is available on the online journal.}}

{\color{Black}{The spectra fitting parameters (temperature, volume emission measure, abundances, energy gain, and offset) were adjusted in OSPEX to improve the goodness of fit  determined by the reduced chi-squared statistic. We define the chi-squared statistic as: 
\begin{equation}
\chi^2_{red}= \sum_{i} \frac{1}{\nu}  \left[ \frac{(O_i - M_i)^2}{\sigma^2_i}  \right]
\end{equation}
where, $i$ is each energy bin, $\nu$ is the number of degrees of freedom, $O$ are the observations, $M$ the fitting model,  $\sigma$ are the uncertainties of the expected count rate assuming Poisson statistics. The instrumental effects are included in the Detector Response Matrix (DRM) which is used to make the model.}}

{\color{Black}{ In our reduced chi-square calculation, each energy bin contains at least 2 counts. Increasing the number to 10 counts per bin does not make a major difference in the estimated parameters. In the appendix, Figure \ref{fig:enhanced_subtracted} compares the spectra fit with a minimum of 10 counts per bin (left column) with a spectra fit with a minimum of 2 counts per bin (middle column). The energy range where the fit is performed is smaller, but the derived temperature and volume emission measure values closely resemble each other. Additionally, the time variation of the reduced chi-squared and average reduced chi-square during the flare are similar in value. We provide a video with Figure \ref{fig:enhanced_subtracted} to highlight the similarities in the spectra.  Moreover, the majority of the bins contributing to the chi-square statistic are at the lower energy range. There are 100 to 1000 times more counts in the lower energy range than in the higher energy range. Thus, in our chi-square calculation, the lower energy bins have a higher contribution to the chi square statistic.}}
 
\begin{figure}[ht!]
\plotone{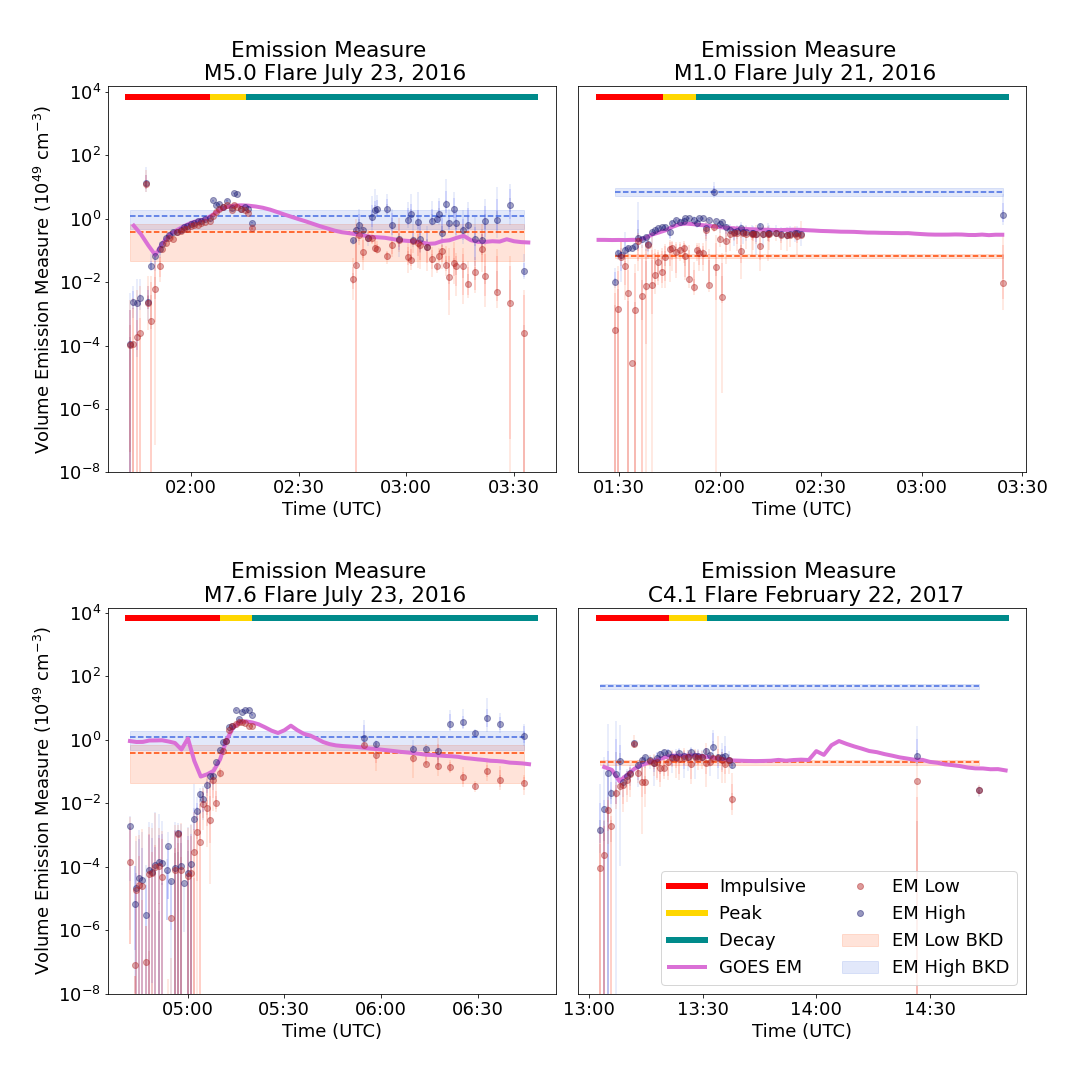}
\caption{Volume emission measure plot for four flares observed by MinXSS-1. The top left is the July 23, 2016 M5.0 flare, top right the July 21, 2016 M1.0 flare, bottom left is the July 23, 2016 M7.6 and bottom right is the C4.1 February 22, 2017. These plots show the two background volume emission measures used in red and blue horizontal dashed lines with its respective shaded uncertainty. The best fit for the two volume emission measure components during the flare are denoted in circles. The uncertainties shown for the best fit parameters are calculated using the Monte Carlo analysis available in OSPEX. Additionally, we plot the derived GOES volume emission measures available in SolarSoftware. We note that the best fit volume emission measures closely follow the GOES derived volume emission measures. \label{fig:VEM_Minxss}}
\end{figure}

\begin{figure}[ht!]
\plotone{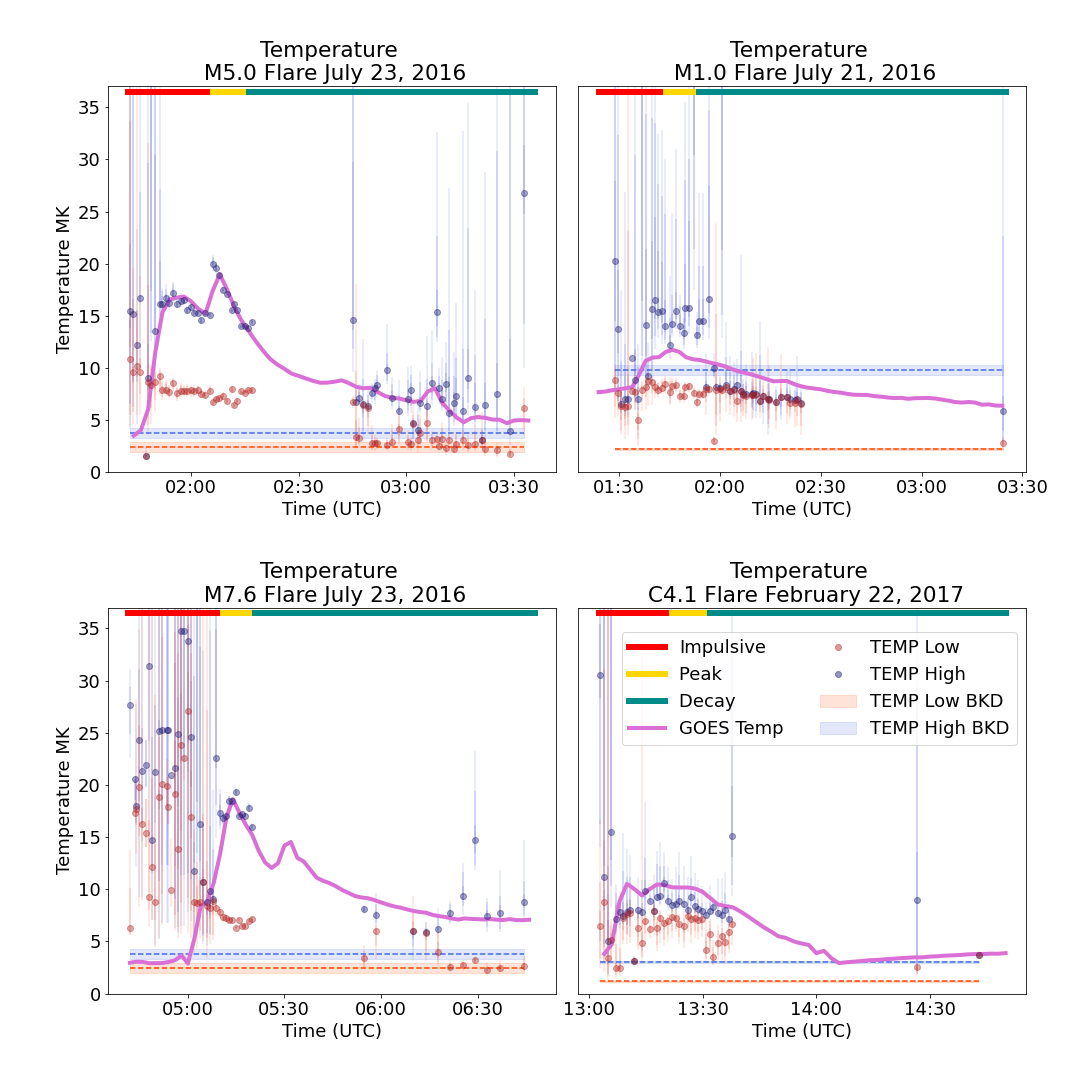}
\caption{Temperature plots for four flares observed by MinXSS-1.  The top left is the July 23, 2016 M5.0 flare, top right the July 21, 2016 M1.0 flare, bottom left is the July 23, 2016 M7.6 and bottom right is the C4.1 February 22, 2017. These plots show the two background temperatures used in red and blue horizontal dashed lines with its respective shaded uncertainty. The best fit for the two temperature components during the flare are denoted in circles. The uncertainties shown for the best fit parameters are calculated using the Monte Carlo analysis available in OSPEX. We note that the uncertainties during the impulsive phase are large due to the lack of statistical significant signal.   \label{fig:Temp_Minxss}}
\end{figure}

The derived two volume emission measures and two temperatures of the four flares are shown in Figures \ref{fig:VEM_Minxss} and \ref{fig:Temp_Minxss}. The two background volume emission measures and temperature are denoted in red and blue horizontal dashed lines with corresponding uncertainties shaded. The best fit for the two value volume emission measures and temperature parameters during the flare are denoted in circles. The uncertainties shown for these four flares are calculated using the Monte Carlo analysis available in OSPEX (\cite{2020TolbertSchwartzOSPEX}, \url{https://hesperia.gsfc.nasa.gov/ssw/packages/spex/doc/ospex_explanation.htm#OSPEX\%20Monte\%20Carlo\%20Analysis}). The uncertainties of all other flares in this study are the basic OSPEX returned fits uncertainties, assuming a local Gaussian shape. Both volume emission measures and temperatures of these flares rise with SXR and show maxima near the SXR peak of each flare. Additionally, we plot the GOES/XRS volume emission measures and temperatures derived from the ratio of the two GOES/XRS channels available in SolarSoftware \citep{white2005updatedgoes, garcia1994temperaturegoes}. We note that the best fit parameters of volume emission measures and temperatures closely follow the GOES derived trends.

MinXSS-1 can derive temperatures in the 1 MK - 30 MK range \citep{moore2018instrumentsminxss}. However, due to the spectral resolution we cannot identify individual lines, but can determine the features of different elements. Depending on the temperature, the elemental abundances are better constrained for some temperature ranges. The higher the flux is, we are able to better determine the temperature and therefore we are able to better constrain the abundances. Moreover, in this study we set a minimal signal-to-noise ratio of nine (S/N = 9) to determine if an elemental abundance parameter is to be fitted. If the count rate flux around the feature of the element  {\color{Black}{has a signal-to-noise ratio greater than 9}}, we autonomously fit the abundance. Otherwise, if {\color{Black}{the signal-to-noise is lower than 9}} we assume the \cite{feldman2003elementalFIP} coronal abundance.

\begin{figure}[ht!]
\includegraphics[scale=0.17]{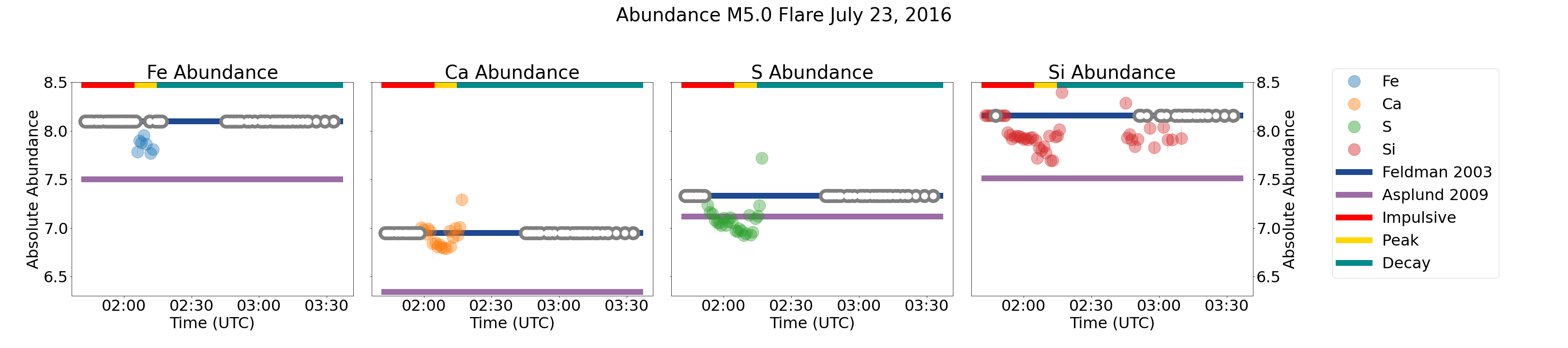}
\includegraphics[scale=0.17]{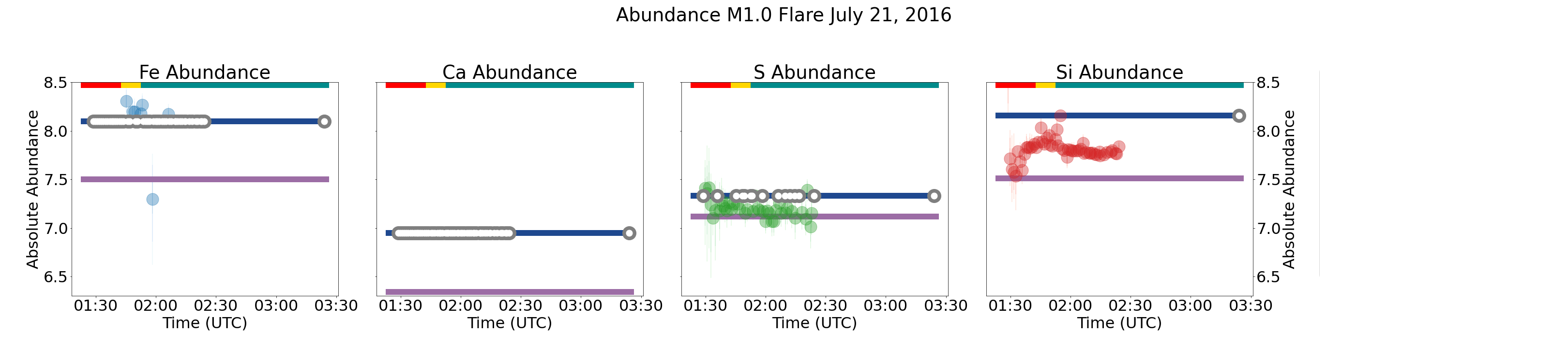}
\includegraphics[scale=0.17]{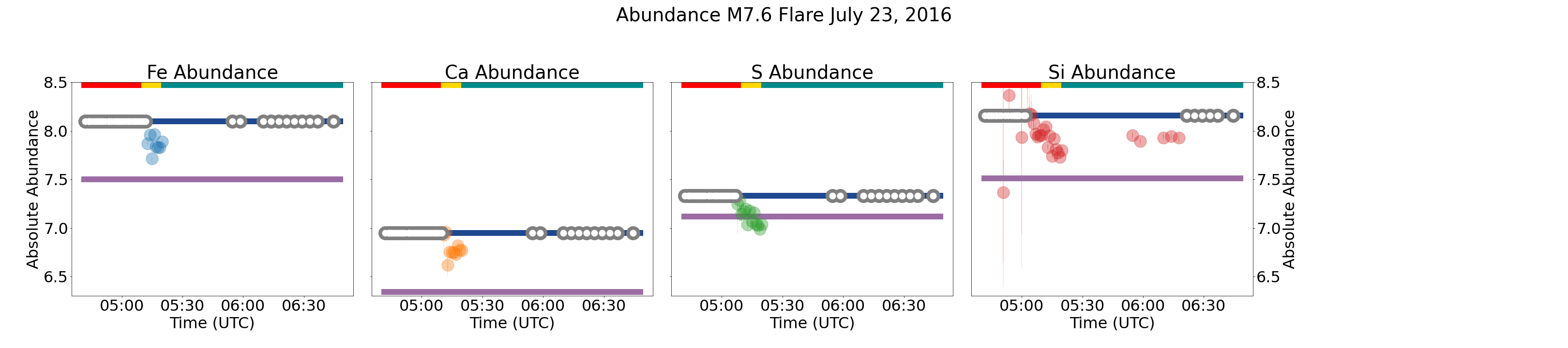}
\includegraphics[scale=0.17]{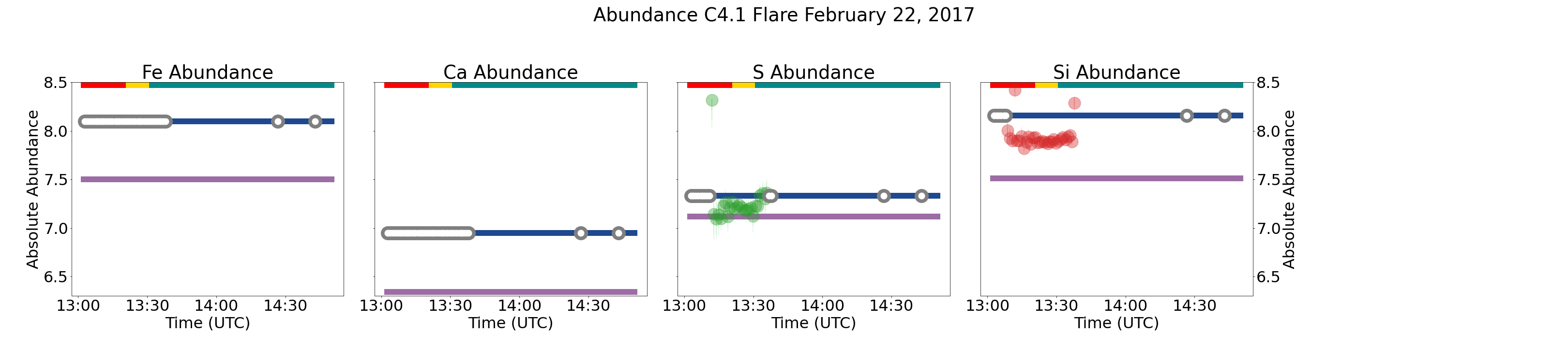}
\caption{Time evolution of absolute abundance for Fe (blue), Ca (orange), S (green), and Si (red) during the four flares observed by MinXSS-1 X123. The navy line shows the coronal abundances from  \cite{feldman2003elementalFIP} and the purple line shows the photospheric abundances from \cite{asplund2009chemical}. The white points are where the model assumes a coronal abundance since the count rate spectra does not have a signal-to-noise ratio for the elemental abundance parameter to be fitted. The absolute abundances are on logarithmic scale with A(H) = 12. We note that at the peak of the flare most elements {\color{Black}{are depleted from coronal towards photospheric values during the peak phase of each flare.}}}
\label{fig:abundance_four_flares}
\end{figure}

{\color{Black}{Figure \ref{fig:abundance_four_flares} shows the temporal evolution in elemental abundance during each of the four flares}}. The white data points denote a low signal-to-noise where the parameters are assumed to be the \cite{feldman2003elementalFIP} coronal values. The fitted parameter with signal-to-noise ratio greater than 9 are shown in light blue (Fe), orange(Ca), green (S), and red (Si). For the Fe abundance we only fit the the feature around 6.7 keV.   {\color{Black}{The FIP multiplier factors from the spectra fit values were converted using the \cite{feldman2003elementalFIP} factors such that: 
\begin{equation}
A(E) = log_{10}(F) + A(E)_{Fl}
\end{equation}

where $A(E)$ is the absolute abundances, $F$ is the FIP multiplier factors, and $A(E)_{Fl}$ are the absolute abundances from \cite{feldman2003elementalFIP}.}} In Figure \ref{fig:abundance_four_flares} {\color{Black}{the}} dark blue line shows the \cite{feldman2003elementalFIP} coronal values and the purple line shows the \cite{asplund2009chemical} photospheric values. During the impulsive phase, the elemental abundances are near coronal levels, but at the peak, the elemental abundances are depleted towards photospheric levels and once again reach coronal levels during the decay phase. This is especially seen with Si - most values are between coronal and photospheric values during the SXR peak.

Figure \ref{fig:all_flares_abund_multi} shows all 21 flares and their averaged absolute abundance of Fe, Ca, Si, and S during the peak of the flare. There is very weak evidence for abundance variations as a function of flare class. We note that absolute abundance of S, and Si at the peak of the flares is lower than the coronal values from \cite{feldman2003elementalFIP}. For some flares greater than M1.0, both Ca and Fe are also lower than coronal abundance. The Fe, Ca abundance variation could be due to larger flares having more flux at the Fe 6.7 keV and Ca 4.0 kev line features and therefore a S/N $>$ 9.

Our findings are consistent with a chomospheric evaporation model. During chromospheric evaporation, the temporal evolution of volume emission measure, temperature, and elemental abundances can be explained. In the impulsive phase of the flare, the magnetic energy is converted to heat from the reconnection site at the apex of the coronal loop, travels down towards the chromosphere and begins evaporating chromospheric plasma into the loop with high velocity \citep{antonucci1985chromosphericEvol}. At this time, both temperature and emission measure increase. As the flare evolves, material from the lower solar atmosphere is transferred into the coronal loops, thereby causing low FIP elemental abundances to fall below coronal abundances. During the peak of the emission, the loop gets filled with the high density evaporated material \citep{ryan2012thermal, klimchuk2017nanoflare}. This evaporated material brings unfractionated chromospheric material into the corona and can explain the abundance depletion from coronal values observed during flares.

\section{Discussion and Conclusion}

In this study, we have performed a spectroscopic analysis of 21 C and M class flares to determine their temporal variation of volume emission measure, temperature, and elemental abundances. This study complements other spectral analysis of other SXR studies. The other studies have estimated flare abundances by various methods, and we compare the results of those studies with the results of this study. Table \ref{tbl:Flare_Params} shows the average absolute abundance of all 21 flares observed by MinXSS-1 at the SXR peak. The weighted arithmetic mean of all flares are: A(Fe) = 7.81, A(Ca) = 6.94, A(S) = 7.28, A(Si) = 7.90 , A(Mg) = 8.34 and, A(Ar) = 6.56. 

The MESSENGER/SAX mission was a broadband X-ray spectrometer that observed the solar X-ray spectra in the range 1.5–8.5 keV \citep{gold2001messenger, santo2001messenger, solomon2001messenger}. \cite{dennis2015solarSAX} analyzed 526 flares from May 2007-August 2013. The authors reported absolute abundances (on a logarithmic scale with H = 12) of A(Fe) = 7.72 $\pm$ 0.11, A(Ca) 6.9 $\pm$ 0.1, A(S) = 7.2 $\pm$ 0.2,  A(Si) = 7.7 $\pm$  0.2, and A(Ar) =  6.8 $\pm$  0.2. The values for Fe, Ca, S, and Si values are within statistical uncertainties of SAX. In \cite{dennis2015solarSAX} all flares analyzed had sufficient intensity to detect emission in the Fe-line complex at 6.7 keV. In our analysis,  flares with a class C4 or greater had sensitivity to observe the Fe-line complex at 6.7 keV.  However, there are only five flares where the Fe 6.7 keV feature is a free parameter in our spectra fits. Therefore, the lack of flares with higher count rate signal at 6.7 keV could lead to some discrepancy between the MinXSS-1 and SAX analysis. 

The X-ray Solar Monitor (XSM) on the Indian lunar space mission Chandrayaan-1 observed the solar 1.8–20 keV X-ray emission \citep{goswami2009chandrayaan1, bhandari2005chandrayaan1}. The mission was only active between November 2008 and August 2009 during a quiet sun, and observed 20 weak A - C class flares. \cite{narendranath2014elementaCH1l} reported a weighted average of A(Fe) = 7.99 $\pm$ 0.09 and  A(Si)= 7.47 $\pm$ 0.1 for a C2.8  flare and for 20 weaker A-C class flares abundances of A(Ca) = 6.6 $\pm$ 0.2,  A(S) = 7.3 $\pm$ 0.3. The value for Fe, Ca, and Si are lower than the results obtained with MinXSS -1. The S  value is within the statistical uncertainties of XSM.

The recent Chandrayaan-2 lunar space mission also has an XSM monitor with an energy range  of 1-15 keV \citep{vadawale2014solarCH2xsm, shanmugam2020solarCH2xsm}. The mission is currently active, but first results of 9 B class flares during the 2020 solar minimum have been reported by \cite{mondal2021evolutionCH2}. The average abundance of these flares were A(S) = 7.0 and A(Si) = 7.69. These values are lower than the results obtained with MinXSS -1. 

\cite{phillips2012rhessi} analyzed the Fe 6.67 keV and Fe+Ni 8 keV emission line complexes in
RHESSI X-ray spectra during 20 solar flares from 2002-2005. In the study, the estimated Fe abundance is A(Fe) = 7.91 $\pm$ 0.10. The Fe abundance estimated with MinXSS-1 is within the statistical uncertainty of RHESSI.

The REntgenovsky Spektrometr s Izognutymi Kristalami (RESIK) crystal spectrometer on the CORONAS-F spacecraft with spectral range 3.3–6.1 Å observed 33 flares from 2002-2003 and derived abundances of A(Si) = $7.53\pm 0.08$, A(S) = $6.91\pm 0.07$, A(Ar) = $6.47\pm 0.07$ \citep{sylwester2015resik}.  These results are all lower than the values obtained with MinXSS -1.  

The Columbia Astrophysics Laboratory graphite crystal spectrometer on the Orbiting Solar Observatory-8 (OSO-8) observed 10 events from 1975-1976 and derived abundances of A(Ca) = $6.5 ^{+0.1}_{-0.2}$, A(Si) = $7.7 ^{+0.2}_{-0.3}$, A(S) = $6.9 ^{+0.1}_{-0.3}$, A(Ar) = $6.4 ^{+0.2}_{-0.3}$ \citep{veck1981abundancesOSO}. Both Ca and S lower than the values obtained with MinXSS-1. Si and Ar are both within the statistical uncertainty of OSO-8. 

{ {\color{Black}{The Solar Dynamics Observatory / EUV Variability Experiment(SDO/EVE) observes both high temperature Fe emission lines (Fe {\footnotesize XV} – Fe {\footnotesize XXIV}) and continuum emission from thermal bremsstrahlung \citep{pesnell2012SDO,warren2014measurementsEVE}. \cite{warren2014measurementsEVE} used a Differential Emission Measure (DEM) analysis on 21 flares from 2011-2013 and estimated an A(Fe) = 7.57, lower than the value obtained with MinXSS-1. }}

When comparing the other elemental abundances of other instruments with MinXSS-1 X123 agrees with the  abundance values of Fe with MESSENGER/SAX and RHESSI, Ca with Chandrayaan-1/XSM, S with MESSENGER/SAX, and Chandrayaan-1/XSM, and Si and Ar with OSO-8. The absolute abundance values of the different elements can differ from instrument to instrument; for MinXSS-1 X123 we have moderate spectra resolution however, other instruments like broad crystal spectrometers have higher spectra resolution and are able to resolve more complex spectra features.

The left panel on Figure \ref{fig:all_instr_flares} compares the absolute abundance obtained with MinXSS-1 and those from MESSENGER/SAX,  Chandrayaan-1/XSM, Chandrayaan-2/XSM, RHESSI, RESIK, OSO-8, SDO/EVE. The right panel on Figure \ref{fig:all_instr_flares} compares the abundances in terms of FIP bias with respect to \cite{asplund2009chemical}. The abundances of Ca, Fe, and Si, were found to have a higher abundance than in the photosphere by a factor of  $\approx$  2-3 , while S and Ar  were found to have a higher abundance factor of $\approx$ 1.5. 

The overall consistency in the MinXSS-1 derived values showing a depletion pattern from coronal values towards photospheric values, conceives the possibility of chromospheric evaporation happening during solar flares. This means that the newly evaporated chromospheric material, from a region beneath the fractionation layer, having almost nonfractionated photospheric plasma, is injected into the coronal loop.

Future flare observation and instruments with better spectral resolution and broad energy ranges can provide new data to explore and constrain elemental abundance. Moreover, having better time cadence can provide information on how quickly the coronal depletion and recovery happen. Future instruments will aid our understanding of the different processes that dominate when plasma is being injected into the coronal loops.

\begin{figure}[ht!]
\includegraphics[scale=0.3]{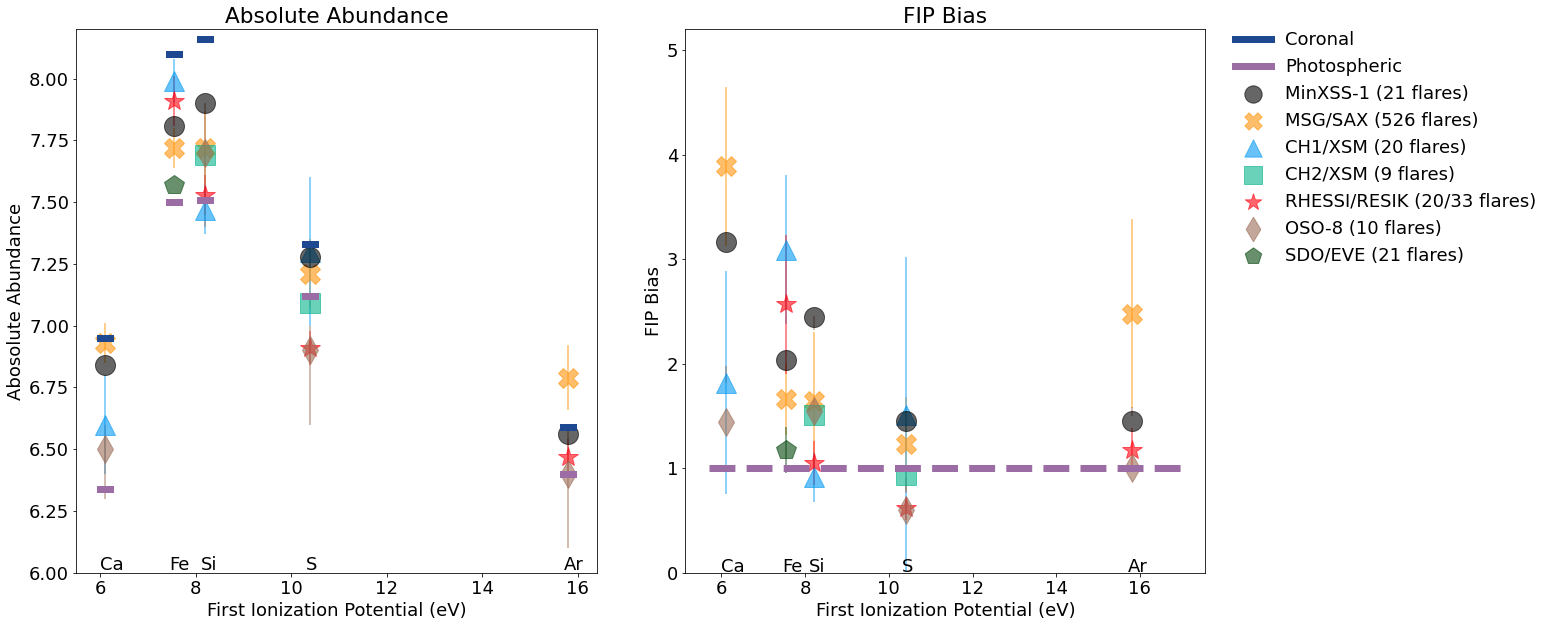}
\caption{ {\color{Black}{Abundance of Fe (7.9 eV), Ca (6.1 eV), S (10.4 eV), Si (8.2 eV) and Ar (15.8 eV) as a function of FIP energy. Black circles are MinXSS-1 weighted arithmetic mean values during the peak of the flare, the orange crosses are the values reported in \cite{dennis2015solarSAX} from 526 flares observed by Messenger/SAX, the light blue triangles are from \cite{narendranath2014elementaCH1l} from 20 flares observed by Chandrayaan-1/XSM, the green squares are from from \cite{mondal2021evolutionCH2} from nine flares observed by Chandrayaan-2/XSM, the red stars are from 20 flares observed by RHESSI \citep{phillips2012rhessi} or 33 flares observed by RESIK \citep{sylwester2015resik}, brown diamonds are from 10 flares observed by OSO-8 \citep{veck1981abundancesOSO}, and green hexagons are from 21 flares observed by SDO/EVE \citep{warren2014measurementsEVE}. The left panel shows the absolute abundances are on logarithmic scale with A(H) = 12. The navy line shows the coronal abundances from \cite{feldman2003elementalFIP} and the purple line shows the photospheric abundances from \cite{asplund2009chemical}. We note how the MinXSS-1 estimated elemental abundances are between the coronal and photospheric values. The right panel shows the elemental abundance in terms of FIP Bias with respect to the \cite{asplund2009chemical} values. The MinXSS-1 estimated low FIP elements(Ca, Fe, Si) are $\approx$ 2-4 times more abundant compare to the photosphere abundance, the mid FIP element (S) and high FIP element (Ar) are $\approx$ 1.5 more abundant compare to the photosphere abundance. }}}  
 \label{fig:all_instr_flares}
\end{figure}


\section{Summary}
In this paper, we present the analysis of 21 well-observed MinXSS-1 solar flares during 2016-2017. We report the temporal evolution of elemental abundances during solar flares. By forward modeling the SXR spectra we obtained parameters of temperature, volume emission measure, and abundances of Fe, Ca, Si, S, Mg, and Ar.  We show that the abundances of these elements are {\color{Black}{depleted from coronal towards photospheric values during the peak phase of each flare.}} This suggests that during the flares, the coronal loops are quickly filled with plasma originating from the lower parts of the solar atmosphere due to chromospheric evaporation. Future SXR spectra observations like the Dual-zone Aperture X-ray Solar Spectrometer (DAXSS) \citep{schwab2020daxss} are expected to provide important constraints to solar corona models.

\begin{acknowledgments}
The authors gratefully acknowledges the guidance and mentorship of Dr. Kelly Holley-Bockelmann. 
The authors acknowledge the support from NASA Heliosphysics Supporting Research Grant (NASA-HSR 16611153), Vanderbilt Bridge Program PhD Fellowship, and the Latino Initiatives administered by the Smithsonian Latino Center. The MinXSS-1 CubeSat mission was supported by NASA Grant NNX14AN84G. We thank the anonymous reviewers for their input and feedback. 
\end{acknowledgments}

\appendix

In this section we explore differences and similarities in the July 23, 2016 M5.0 enhanced spectra, background subtracted spectra, and the results where we consider 2 and 10 counts per energy bin. In the Observations and Data Analysis section, we use a flare enhancement method. With this method, we consider a two thermal spectrum fit for the non-flaring MinXSS-1 spectra. This non-flaring spectrum fit is used as an enhancement when fitting the spectra of the flare. During the observed flare spectra there are possible contributions from non-flaring plasma. When we compared the flare enhanced method and a background subtracted flare spectra, our results were more consistent with the enhanced method. Figure \ref{fig:enhanced_subtracted} middle column shows the spectra fit results of the enhanced method and the right column shows the background subtracted method. We note how the  estimated two volume emission measures and temperatures of the enhanced method are more consistent with the GOES derived values available in Solar Software \citep{white2005updatedgoes,garcia1994temperaturegoes}. Moreover, the enhanced spectra method had a more consistent temporal evolution of chi-square with an approximate value equal to 1. The reduced chi-square in the enhanced method at the SXR peak is 2.29 with an overall global reduced chi-square of 1.62. These values are lower than the background subtracted chi-square values. The corresponding values at the SXR peak with the background subtracted method result in a reduced chi-square of 19.80 and global reduced chi-square of 6.89.

Additionally,  we compared the enhanced spectra fits with different minimum counts per energy bin. On the left column of Figure \ref{fig:enhanced_subtracted}, we show the spectra fit where we consider an average of 10 counts per energy bin.  The middle column shows the spectra fit where we consider an average of 2 counts per energy bin. We note that the high energy limit changes from about 10 keV to about 8 keV when we increase the average number of counts. However, the resulting volume emission measures and temperatures are similar and consistent in both cases. Moreover, the majority of the bins contributing to the chi-square statistic are at the lower energy range, where there are 100 to 1000 times more counts in the lower energy range than in the higher energy.

\begin{figure}[ht!]
\begin{interactive}{animation}{Videos/M5int_time_enhanced_bkd_sub_v3_gain_offset_final.mp4}
\plotone{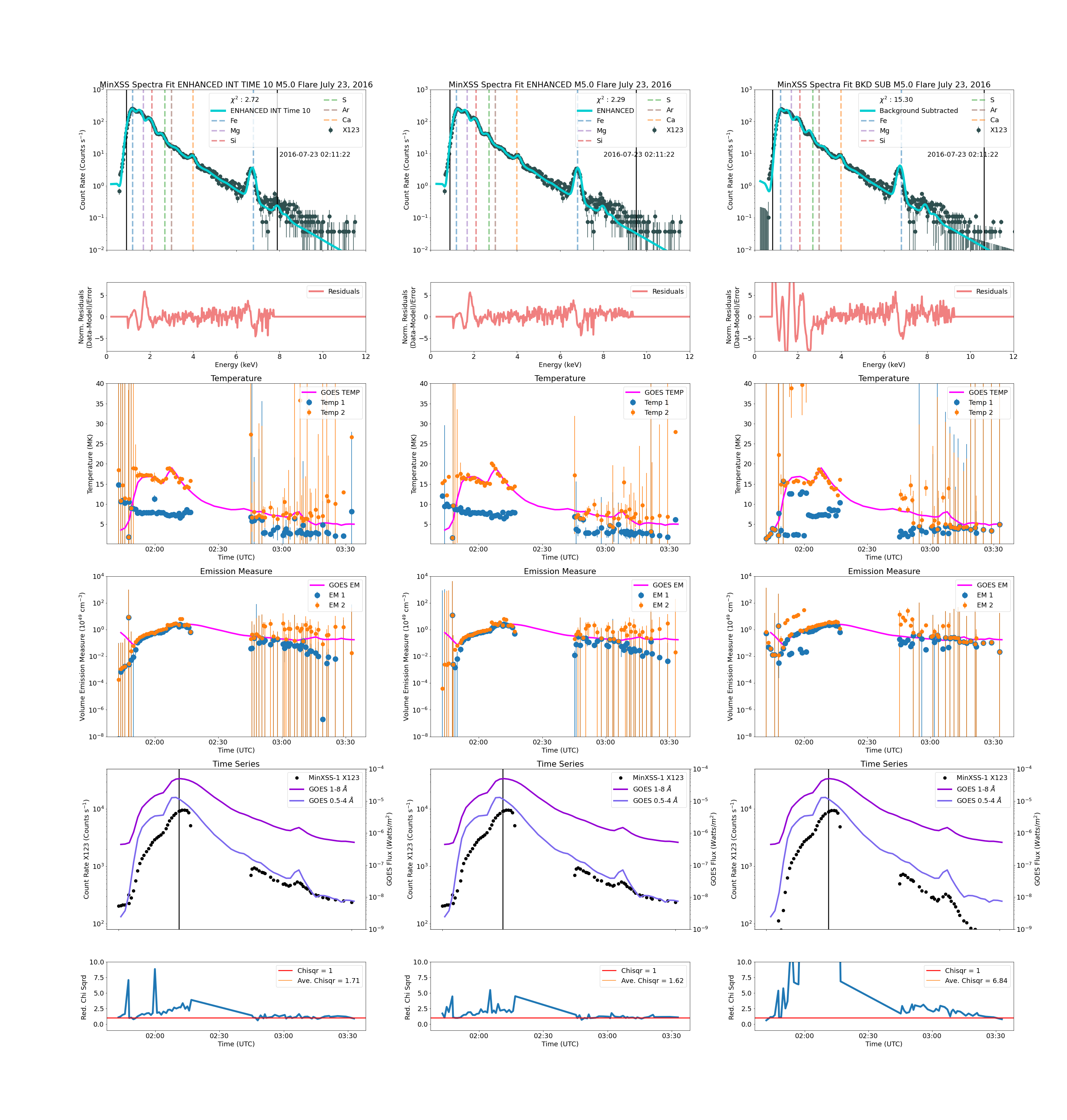}
\end{interactive}
\caption{{Different types of spectra fits for the July 23, 2016 M5.0 flare. On the first column, the spectra fit has at least 10 counts per energy bin. On the second column, the spectra fit is the same as in Figure \ref{fig:Spectra_Minxss}. On the third column, the spectra is background subtracted. In each column below each spectrum are the derived temperatures, the derived volume emission measures, time series and temporal evolution of the reduced chi-squared. The first and second columns have similar temperatures and emission measures with only a few intervals where the reduced chi-square value is different. In the third column the temperatures and volume emission measure have higher values at the impulsive phase of the flare and the values of reduced chi-square are higher than the values from the second column. A video showing the temporal evolution of the spectra fit, and corresponding values of temperature, volume emission measures and reduced chi-square is available in the online version of this paper. At the different phases of the flare, the spectrum changes with respect to changes in temperature and volume emission measure derived values. (Video duration: 00:05).}}
\label{fig:enhanced_subtracted}
\end{figure}

%

\vspace{5mm}
\facilities{MinXSS-1, SDO/AIA (94 \AA), GOES/XRS }


\software{sunpy \citep{sunpy_community2020},  
            OSPEX \citep{2020TolbertSchwartzOSPEX}
            }







\bibliography{sample631}{}
\bibliographystyle{aasjournal}

\end{document}